\documentclass[final,5p,times,twocolumn]{elsarticle}
\biboptions{numbers,sort&compress}
\usepackage{amssymb}
\usepackage{lipsum}
\usepackage{amsmath}
\usepackage{amssymb}
\usepackage{graphicx}
\usepackage{dcolumn}
\usepackage{bm}
\usepackage{braket}
\usepackage{mathrsfs}
\usepackage{booktabs}
\usepackage{multirow}
\usepackage{url}
\usepackage{physics}
\usepackage{subfigure}
\usepackage[colorlinks=true,linkcolor=blue, filecolor=blue,urlcolor=blue,citecolor=blue]{hyperref}
\usepackage{color}
\usepackage{cuted}

\begin{document}
	
	\begin{frontmatter}
% Solving Nuclear resonant states with an HHL-Inspired Quantum Algorithm
% 如果站在郭老师的角度也许需要一篇 CMR for resonance
% 站在konig的角度则是EC
% QNN+IACCC的文章也许直接换成toy model更好一些吗？ 还是说就按cluster model 来？
% 第二步需要考虑的问题是必须要在量子设备上跑起来，这一点是目前的弱点，容易被别人针对
% 计划表： 审稿意见先稍微放一下，把上面这几个点都做一做试试看
% 也许CMR和CSM可以放在一起做，或者说就干脆分开
%CMR/CSM+HHL的优先级目前判断是高于HHL+R矩阵的，且稍高于QNN，因此我们现在不妨就用模拟的HHL先把CSM和CMR搞定，且相比于CSM而言CMR本身是有一定优势的，宽束缚态的稳定性以及contour的选择比CSM的角度确定感觉可能更好一点
% 也许CMR和CSM放一块也可行
% 另外一点就是站在郭老师的角度来说，相对论的狄拉克方程也是能拿到credit的点之一，因此这个部分也需要布局一下，目前想到的三个点就是CSM，CMR以及狄拉克方程
% 至于要分几篇来写这里先不着急决定
% 目前眼下最重要的任务就是CSM+HHL（CMR+HHL的也一块考量一下）	
% 然后优先级稍低的就是QNN，以及QC+R矩阵
% 最后可以想办法把plb里面的比特数再推广一下看能不能有一些新内容

% 现在的目标就是
\title
%{A new Quantum Algorithm for Solving resonant states in Nuclear Systems}
%{Harrow-Hassidim-Lloyd based quantum algorithm for solving resonances with eigenvector continuation}
{Iterative Harrow-Hassidim-Lloyd quantum algorithm for solving resonances with eigenvector continuation}

\author[first]{Hantao Zhang}
\author[second,fourth]{Dong Bai}
%\author[first]{Zhen Wang}
\author[first,third]{Zhongzhou Ren\corref{cor1}}
\ead{zren@tongji.edu.cn}
\cortext[cor1]{Corresponding author}

\affiliation[first]{organization={School of Physics Science and Engineering},%Department and Organization
	addressline={Tongji University}, 
	city={Shanghai},
	postcode={200092}, 
	% state={},
	country={China}}
\affiliation[second]{organization={College of Mechanics and Engineering Science},%Department and Organization
	addressline={Hohai University}, 
	city={Nanjing},
	postcode={211100}, 
	% state={},
	country={China}}
\affiliation[third]{organization={Key Laboratory of Advanced Micro-Structure Materials},%Department and Organization
	addressline={Tongji University}, 
	city={Shanghai},
	postcode={200092}, 
	% state={},
	country={China}}
\affiliation[fourth]{organization={Shanghai Research Center for Theoretical Nuclear Physics},%Department and Organization
	addressline={NSFC and Fudan University}, 
	city={Shanghai},
	postcode={200438}, 
	% state={},
	country={China}}

%\author{Hantao Zhang}
%\email{zhang\_hantao@foxmail.com}
%\affiliation{School of Physics Science and Engineering, Tongji University, Shanghai 200092, phina}
%
%\author{Dong Bai}
%\email{dbai@hhu.edu.cn}
%\affiliation{College of Science, Hohai University, Nanjing 211100, Jiangsu, phina}
%
%\author{Zhen Wang}
%\email{wang\_zhen@tongji.edu.cn}
%\affiliation{School of Physics Science and Engineering, Tongji University, Shanghai 200092, phina}
%
%\author{Zhongzhou Ren}
%\email[Corresponding author: ]{zren@tongji.edu.cn}
%\affiliation{School of Physics Science and Engineering, Tongji University, Shanghai 200092, phina}
%\affiliation{Key Laboratory of Advanced Micro-Structure Materials, Ministry of Education, Shanghai 200092, phina}

\begin{abstract}

We propose a novel quantum algorithm for solving nuclear resonances, which is based on the iterative Harrow-Hassidim-Lloyd algorithm and eigenvector continuation with complex scaling. To validate this approach, we compute the resonant states of $\alpha-\alpha$ system and achieve results in good agreement with traditional methods. Our study offers a new perspective on calculating eigenvalues of non-Hermitian operators and lays some groundwork for further exploration of nuclear resonances using quantum computing.

\end{abstract}


\end{frontmatter}
% 目前的文章中先不使用真机，统一用模拟的方式处理

% 束缚态的话随便找一个例子，最终能求解出基态能量即可
% 控制们的顺序需不需要也搞清楚画个图？
% 需要出现不同学习率的能量收敛曲线

% 散射态的话也随便找一个例子，可以用HHL求矩阵的逆即可
% 这里需要出现HHL的基本电路图和数值结果（可能需要搞清楚函数中integer的含义）

% 共振态的话可以先把基矢量数目放少一些，目前这一部分最拿不准

% 需要再重新梳理一下内容的安排，传统方法作为关键词来安排内容还是说用算法来安排内容，或者用物理态来安排内容？

%% ACCC ；  R矩阵
%% QNN  ；HHL
%% bound； scattering and resonance

%这样梳理会发现第二种和第三种实际上没有区别

%而且HHL来处理非厄米与R矩阵中处理resonance之间是存在递进关系的
% R矩阵中处理resonance不仅是非厄米而且是需要迭代的
% 这么去看的话，如果用CSM那么计算量好像更大
% 而且一旦HHL+非厄米基本成型，那么用在complex potential+trap中也是非常自然的

% 目前感觉ACCC的性价比稍微低一些，可能作为QNN的子标签比较好

% 也许这样比较好，用QNN先来写一个文章，然后再考虑HHL+非厄米
% 有一点很显然，R-matrix+scattering+HHL来做矩阵求逆这个分量不太足

% 目前看起来还是以算法为标签可能更好，这样QNN的文章里就会有R矩阵+bound的内容而且总体看上去比较清晰

%\section{Introduction}
%
%\section{Theoretical Formalism}\label{}
%
%
%\section{Numerical Results}
%
%\section{Conclusions}

%\section{quantum gate}
%NOT gate:X
%
%Y:
%
%Z:
%
%X,Y,Z Pauli matrix
%
%Hadamard gate:
%
%P gate
%
%
%
%CNOT Gate:
\section{introduction}

Quantum computing, a rapidly advancing field at the nexus of quantum mechanics and computer science, has demonstrated significant potential in addressing problems intractable for classical systems. Its applications span diverse scientific domains, including computational chemistry, high energy physics, nuclear physics, and more. At the core of quantum computing's transformative capabilities lie quantum entanglement and quantum information, which provide unparalleled computational efficiency and power. In nuclear physics, the study of entanglement properties in hadrons and nuclei is emerging as a critical area of interest \cite{Savage:2023qop,Klco_2022,Ho_2016,PhysRevD.95.114008,PhysRevD.98.054007,PhysRevLett.122.102001,Tu_2020,BEANE2021168581,ISKANDER2020135948,Kruppa_2021,doi:10.1142/S0217751X21502055,PhysRevD.104.L031503,PhysRevC.106.024303,PhysRevC.103.034325,PhysRevD.104.074014,PhysRevD.106.L031501,Bai:2022hfv,Johnson_2023,EHLERS2023169290,Tichai_2023,Pazy_2023,Bulgac_2023,PhysRevA.105.062449,PhysRevC.107.044318,PhysRevC.105.014307,Bai:2023rkc,Robin_2023,Gu_2023,Sun_2023,Bai:2023tey,Bai:2023hrz,Bai:2024omg,miller2023entanglementmaximizationlowenergyneutronproton,Hengstenberg_2023,P_rez_Obiol_2023,Gorton_2024} , further underscoring the intersection of quantum computing and fundamental science.
For researches in nuclear physics, quantum computing’s powerful parallel processing capabilities provide new pathways for investigating complex quantum mechanical problems. Specifically, quantum computing has shown superiority in solving large-scale many-body systems and handling state evolution in high-dimensional spaces, offering unprecedented opportunities to explore complex nuclear systems. {\color{black}As some ground work, in the study of nuclear reactions, quantum computing and derivative algorithms have already been increasingly successfully applied \cite{Roggero_2019,Mueller_2020,Turro_2023,Baroni_2022,bedaque2022radiativeprocessesquantumcomputer,turro2024evaluationphaseshiftsnonrelativistic,Du_2021}.}

The growth of quantum computing has been driven by the development of crucial algorithms that address challenges beyond classical computing. For instance, algorithms like Harrow-Hassidim-Lloyd (HHL) \cite{PhysRevLett.103.150502} has been employed to solve linear systems, quantum phase estimation (QPE) \cite{kitaev1995quantummeasurementsabelianstabilizer,Nielsen_Chuang_2010}, the variational quantum eigensolver (VQE)  \cite{osti_1623945,Peruzzo_2014,McClean_2016,PhysRevLett.120.210501,RevModPhys.94.015004,PhysRevC.104.024305}, the  {\color{black}variants} of VQE \cite{Higgott_2019,McArdle_2019,Yuan_2019,Grimsley_2019,PhysRevResearch.2.043140,Stokes_2020,Gomes_2021,PRXQuantum.2.020310,PhysRevC.105.064317,Koczor_2022}, and the quantum  neural network (QNN) \cite{Cong_2019,Beer_2020,Abbas_2021,Pan_2023,Jin_2024}  have achieved substantial progress in solving eigenvalue problems in quantum systems, simulating molecular and nuclear structures, and modeling chemical and nuclear reactions. 
%Through these foundational quantum algorithms, researchers can solve specific linear algebra problems, differential equations, and eigenvalue problems in quantum systems within the quantum computing framework. 

 Although significant progress has been made in solving  Hermitian systems, challenges persist when tackling non-Hermitian problems with quantum computing.
As a important topic in non-Hermitian quantum dynamics, the resonant state describes metastable systems with finite lifetimes and  plays a key role in studying scattering and reaction cross sections, nuclear decay, nuclear fusion, and also the properties of nuclear matter etc.. Therefore, developing efficient quantum algorithms with low time complexity for nuclear resonant states remains a critical and unresolved challenge. {\color{black} From  experimental perspective, the significant attenuation of cosmic-ray-induced neutrino and muon fluxes in extraterrestrial environments with negligible atmospheres facilitates advancements in nuclear reaction research \cite{Zhang:2024xrj}. Additionally, investigating the use of quantum computing to simulate the effects of cosmic-ray-induced particles on nuclear experiments may be also an area that warrants further exploration.}

%% 20241.12.31  
%% EC和emulator分开写
%% 还需要想办法把zhangxilin solar的文章引一下
%% 怎么引呢？这个文章是说外太空中有些区域的粒子流强度很小，对研究核反应等的影响较小
%% trap里面还需要加一点参考文献吗？

%% 学习一下Dean Lee的风格

In our previous work, we have conducted some studies on non-Hermitian nuclear resonant states \cite{Zhang:2024rpa}. {\color{black} In this work, we propose another novel quantum algorithm for solving resonances, which is based on the iterative HHL (IHHL) algorithm combined with complex scaling method (CSM) \cite{Aguilar1971ACO,Balslev1971SpectralPO,10.1143/PTP.116.1,Myo:2014ypa,PhysRevC.89.034322,Zhang:2022rfa,Zhang:2023dzn,Myo:2023btg,Zhang:2024ril,Zhang:2024gac,10.1088/1674-1137/ad88fa,10.1088/1674-1137/ad9a8c} and eigenvector continuation (EC) \cite{PhysRevLett.121.032501,PhysRevC.107.064316,PhysRevC.101.041302,KONIG2020135814,COMPANYSFRANZKE2022137101,10.1093/ptep/ptac057,PhysRevLett.126.032501}.  Eigenvector continuation has already  found many applications in low energy nuclear physics and in particular, it has
been utilized to build emulators in the context of nuclear scattering \cite{Melendez_2022,PhysRevC.106.054322,10.3389/fphy.2022.1092931,FURNSTAHL2020135719,DRISCHLER2021136777,PhysRevC.103.014612,PhysRevC.106.024611,PhysRevC.107.054001}.} The HHL algorithm is a pure quantum algorithm for solving linear systems,  which is constructed based on the QPE algorithm and has been successfully applied in various research fields. Considering the non-Hermitian property introduced by CSM, we transform the complex eigenvalue problem into an iterative one and introduce a novel iterative HHL algorithm. 
Additionally, in order to handle resonant states more efficiently, we incorporate eigenvector continuation with complex scaling, which leverages a linear combination of known eigenvectors at controlled parameters to  approximate the target resonant eigenstates. Its advantage lies in significantly reducing computational dimension and  qubits required for quantum computation, thereby enhancing numerical stability and computational efficiency, particularly in handling many-body systems, parameter-dependent problems and so on. To validate the reliability of our approach, we perform quantum simulations to compute the resonant state of the $\alpha-\alpha$ system. Our new algorithm also holds potential for applying the trap method \cite{Zhang:2024vmz,Zhang:2024mot,Zhang:2024ykg} to handle scattering phase shifts in the complex domain.

The rest parts are organized as follows: In Sec.II, we introduce the framework of iterative Harrow-Hassidim-Lloyd algorithm and eigenvector continuation with complex scaling. In Sec.III, the numerical results  of resonance in $\alpha-\alpha$ system are presented and discussed. Sec.IV summarizes the article. Some numerical details are listed in the appendix.

%具体来说通过一般的eigenvector continuation我们可以得到某些参数对应的基态束缚态，然后借助complex scaling 我们可以完成bound to resonance的计算任务。这一部分基态束缚态的计算我们可以通过QNN，VQE等量子算法完成，而涉及到complex scaling的非厄米算符本征值问题我们提出了一种基于HHL算法的复数本征值求解器，可以得到非厄米矩阵的全部特征值和特征向量。
\section{theoretical formalism}
\label{Theoretical Formalism}
\begin{figure*}[htbp] 
	\centering
	{\includegraphics[width=0.85\textwidth]{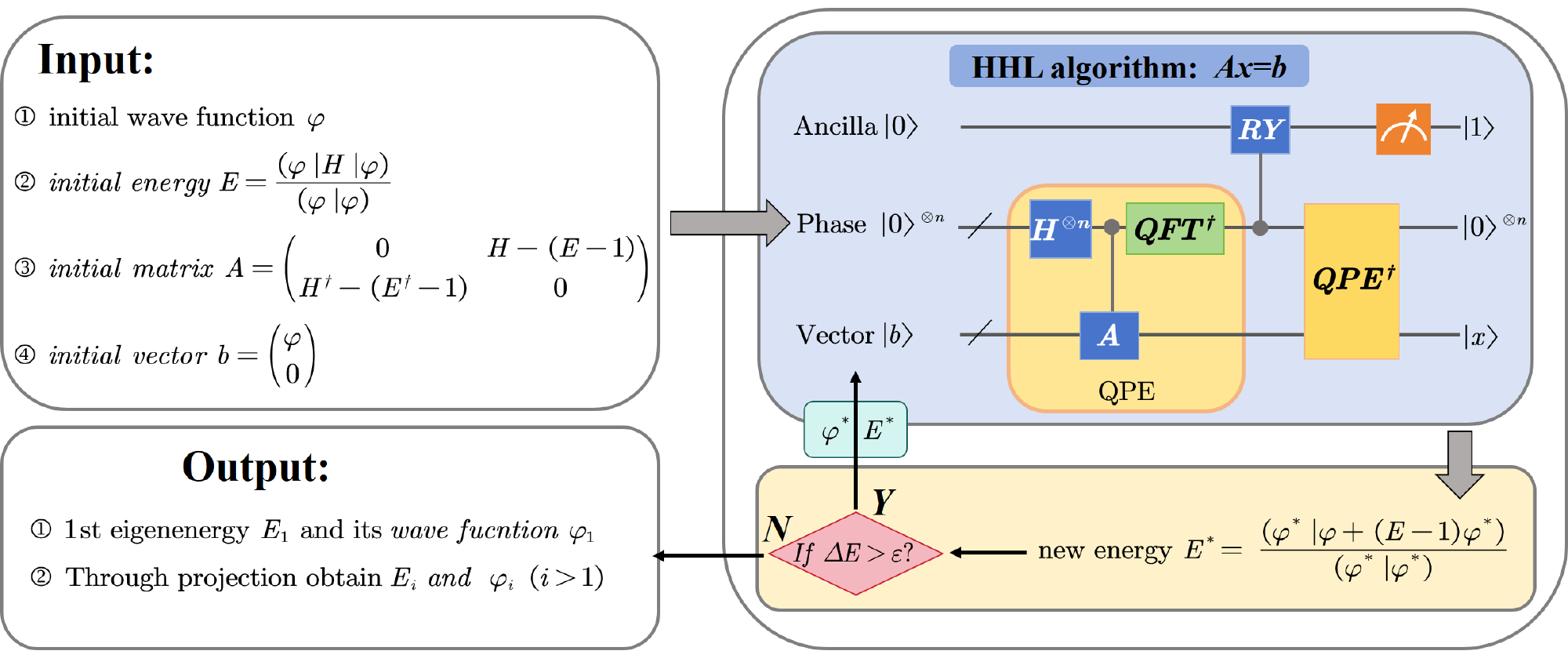} }
	\caption{
		Schematic diagram of iterative resonant state solution based on the HHL algorithm. Input part: Given an initial trial wave function $\phi$ and calculate the corresponding energy expectation value $E$. 
		Then, expand the original non-Hermitian matrix into a Hermitian matrix $A$  and also expand the initial wave function $\phi$ into a vector $b$ as the initial inputs for the HHL algorithm. After solving the linear equation 
		$Ax=b$ using the HHL algorithm, the new wave function $\phi^{*}$ is extracted from $x$ to calculate the new energy expectation value $E^{*}$. 
		If the energy does not meet the specified accuracy $\varepsilon$, then we need to construct a new linear equation using the new wave function $\phi^{*}$ and the new energy $E^{*}$, and continue solving with the HHL algorithm until the specified accuracy is achieved. Finally, output the first convergent eigenvector $\phi_1$ and eigenenergy $E_1$. Consequently, by employing a projection method, we can enforce the orthogonality of the wave function to the previously solved ones and continue to obtain the remaining eigenvectors using the iterative HHL algorithm.}	\label{HHL_non_hermitian} 
\end{figure*}

\subsection{Iterative Harrow-Hassidim-Lloyd (IHHL) algorithm}

When selecting a suitable set of orthonormal basis functions $\{\phi_i\}$, the parameterized Hamiltonian $H(\lambda)$ can be expressed in the following form,
\begin{equation}
	\begin{aligned}\label{}
		&H(\lambda)=\sum_{i,j}^{N}h_{i,j}(\lambda)a_i^{\dagger}a_j^{},\ \ \ h_{i,j}(\lambda)=\bra{\phi_i}|{T+V_N(\lambda)+V_C}\ket{\phi_j }.
	\end{aligned}
\end{equation}

%\begin{equation}
%	\begin{aligned}\label{}
%%		&h_{i,j}=\int \phi_i^{}(r)(T+V(r))\phi_j(r)d^3r,
%		&h_{i,j}=\bra{\phi}{T+V}\ket{\phi}
%	\end{aligned}
%\end{equation}

The fermionic creation and annihilation operators can be expressed through the Jordan-Wigner transformation \cite{Jordan1928berDP},
\begin{equation}
	\begin{aligned}\label{}
		&a_j^{\dagger}=\dfrac{1}{2}(X_j-iY_j)\otimes Z_{j-1}^{\mathscr{D}},\\
		&a_j^{}=\dfrac{1}{2}(X_j+iY_j)\otimes Z_{j-1}^{\mathscr{D}},
	\end{aligned}
\end{equation}
where $X$, $Y$, and $Z$ are Pauli operators, $Z_{j-1}^{\mathscr{D}}$  is defined as
\begin{equation}
	\begin{aligned}\label{}
		&Z_{j-1}^{\mathscr{D}}=Z_{j-1}\otimes Z_{j-2}\otimes \cdots \otimes Z_0.
	\end{aligned}
\end{equation}

In addition to the Jordan-Wigner transformation, there are other quantum mapping methods. For example, for $2^{m}\times2^{m}$ matrices, one can choose to represent the Hamiltonian matrix using $m$ qubits, which can effectively reduce the number of required qubits,

\begin{equation}
	\begin{aligned}\label{}
		&H(\lambda)=\sum_{i_1,i_2,\cdots,i_m=0,1,2,3}c_{i_1,i_2,\cdots,i_m}(\lambda)(\sigma_{i_1}\otimes\sigma_{i_2}\otimes\cdots\otimes\sigma_{i_m}),
	\end{aligned}
\end{equation}
where
\begin{equation}
	\begin{aligned}\label{}
		&c_{i_1,i_2,\cdots,i_m}(\lambda)=\dfrac{1}{2^{m}}Tr((\sigma_{i_1}\otimes\sigma_{i_2}\otimes\cdots\otimes\sigma_{i_m})H(\lambda)), \ \sigma=\{I,X,Y,Z\}.
	\end{aligned}
\end{equation}

The Harrow-Hassidim-Lloyd algorithm is a quantum algorithm designed to solve  linear equation $Ax=b$, where $A$ is a Hermitian operator. As displayed in Fig.1, it leverages quantum phase estimation (QPE) to find the eigenvalues of the matrix $A$, then uses quantum gates to compute the inverse of these eigenvalues. Finally, the algorithm reconstructs the solution vector $x$ in quantum form. HHL algorithm can provide exponential speedup over classical methods for Hermitian matrices. 
%% 这里不够需要再写一些

According to the ABC theorem \cite{Aguilar1971ACO,Balslev1971SpectralPO}, the  original Hamiltonian $H({\lambda})$ will be transformed into a complex scaled one, $H^{\theta}(\lambda)=e^{-2i\theta}T+V_N(\lambda,re^{i\theta})+e^{-i\theta}V_C$. Therefore, to utilize the HHL algorithm to solve resonances,  a non-Hermitian complex scaled Hamiltonian matrix $H$ should be extended to be a larger Hermitian matrix $A$, 
\begin{equation}
	\begin{aligned}\label{}
		&A=
		\begin{pmatrix}
		    0& H \\
		    H^{\dagger}&0	
	    \end{pmatrix}
    .
	\end{aligned}
\end{equation}

Additionally, to address the eigenvalue problem for resonant states, we introduce an iterative method, specifically we can reformulate the Schrödinger equation, so that the final eigenvector becomes the fixed point of the equation as follows,
\begin{equation}
	\begin{aligned}\label{}
		&C(E,\beta)\phi=\phi,
	\end{aligned}
\end{equation}
where  matrix elements of $C$ operator is defined by
\begin{equation}
	\begin{aligned}\label{}
		&C_{ij}(E,\beta)=(\phi|\dfrac{e^{-2i\theta}T+V_N(\lambda,re^{i\theta})+e^{-i\theta}V_C-(E-\beta)}{\beta}|\phi),\ \ \  \beta\neq0.
	\end{aligned}
\end{equation}
where parentheses $"()"$ denotes the $c$-product.

If we simply let $\beta$ equal to 1, then larger Hermitian matrix $A$ can be written as,
\begin{equation}
	\begin{aligned}\label{LargerM}
		&A=
		\begin{pmatrix}
			0& C(E,1) \\
			C^{\dagger}(E,1)&0	
		\end{pmatrix}
	.
	\end{aligned}
\end{equation}

After the above processing, we can utilize the HHL algorithm to iteratively solve for the eigenenergies of resonant state. A more detailed algorithmic diagram is presented in Fig. 1. Our algorithm can be mainly divided into three parts: input, the self-consistent iterative HHL quantum algorithm, and output.  At first, we define an initial trial wave function $\phi$ to construct vector $b$ and input it into the HHL algorithm. Next, the non-Hermitian Hamiltonian is extended into a Hermitian operator $A$ as in Eq.(\ref{LargerM}), where $E$ is the initial energy corresponding to the initial wave function $\phi$, which can be either real or complex. In addition, it is important to note that generally the input vectors for the HHL algorithm should be real numbers. Therefore, when dealing with complex wave functions, we need to separate the real and imaginary parts and apply the HHL algorithm to each part individually.

The HHL algorithm solves for the unknown vector $x$, namely corresponding to the new complex wave function $\phi^{*}$. The $c$-product of the new and old wave functions is then used to calculate the new complex energy $E^{*}$ as shown in Fig.1. By setting a error tolerance $\varepsilon$, we can determine whether to continue iterating with the HHL algorithm until the desired accuracy is reached, and eventually output the convergent eigenvalue and eigenvector of the complex scaled Hamiltonian. At this stage, one of the $N$ eigenvectors is obtained. To find the remaining eigenvectors, the projection method can be applied. Specifically, during HHL iterations, the target eigenvector is made orthogonal to the previously obtained eigenvectors. 

%To reduce computation, orthogonalization can be performed only on the initial input vector, though this is not always effective.

\subsection{Eigenvector continuation}

Eigenvector continuation is an approach that can extract the eigenstates $\phi(\lambda_{\odot})$ of a Hamiltonian $H(\lambda_{\odot})$ with target  parameter $\lambda_{\odot}$ by obtaining eigenstates $\phi_T=\{\phi(\lambda_{i})\}$ of the Hamiltonian $\{H(\lambda_i)\}$. Explicitly speaking,  the parameter set $\{\lambda_i\}$ is served as training data and the corresponding training eigenvectors will be utilized to form a new basis to solve the eigenvalues and eigenvectors of target Hamiltonian $H(\lambda_{\odot})$.
Therefore eigenvector continuation can generally reduces the dimension of the eigenvalue problem from a large Hilbert
space to a smaller subspace spanned by the training eigenvectors $\phi_T$. In this way the computational cost
for each target parameter can be significantly reduced.

One of the  fundamental applications of eigenvector continuation is to estimate bound states from bound states.  Specifically, for extracting bound states the generalized eigenvalue
problem should be solved,
\begin{equation}
	\begin{aligned}\label{}
		&H^{EC}\ket{\phi(\lambda_\odot)}=E(\lambda_{\odot})N^{EC}\ket{\phi(\lambda_\odot)},
	\end{aligned}
\end{equation}
where the Hamiltonian and overlap matrix elements $H(\lambda_{\odot})^{EC}$, $N(\lambda_{\odot})^{EC}$ are constructed by training eigenvectors of bound states, respectively,
\begin{equation}
	\begin{aligned}\label{}
		&H^{EC}_{ij}=\bra{\phi(\lambda_{i})}{H(\lambda_{\odot})}\ket{\phi(\lambda_{j})},\ \ \ N^{EC}_{ij}=\bra{\phi(\lambda_{i})}\ket{\phi(\lambda_{j})}.
	\end{aligned}
\end{equation}

In addition, by introducing complex scaling $r\rightarrow re^{i\theta}$ (or equivalent $k\rightarrow ke^{-i\theta}$), EC can also be extended to perform the task of estimating resonant states from bound states. It is noticeable that 
the inner product should be replaced by the $c$-product,
\begin{equation}
	\begin{aligned}\label{cproduct}
		&H^{EC,\theta}_{ij}=(\phi(\lambda_{i})|H^{\theta}(\lambda_{\odot})|\phi(\lambda_{j})),\ \ \ N^{EC}_{ij}=(\phi(\lambda_{i})|\phi(\lambda_{j})).
	\end{aligned}
\end{equation}

Alternatively, we can use complex scaled Hamiltonian and real training eigenvectors as in Eq.(\ref{cproduct}) or complex scaled training eigenvectors and real Hamiltonian to extract resonant energy. It is important to note that in quantum computing, an orthonormal basis is typically required. Therefore, after obtaining the training eigenvectors for bound states using eigenvector continuation, Schmidt orthogonalization should be employed.
 
  \begin{figure}[htbp] 
 	\centering
 	{\includegraphics[width=0.5\textwidth]{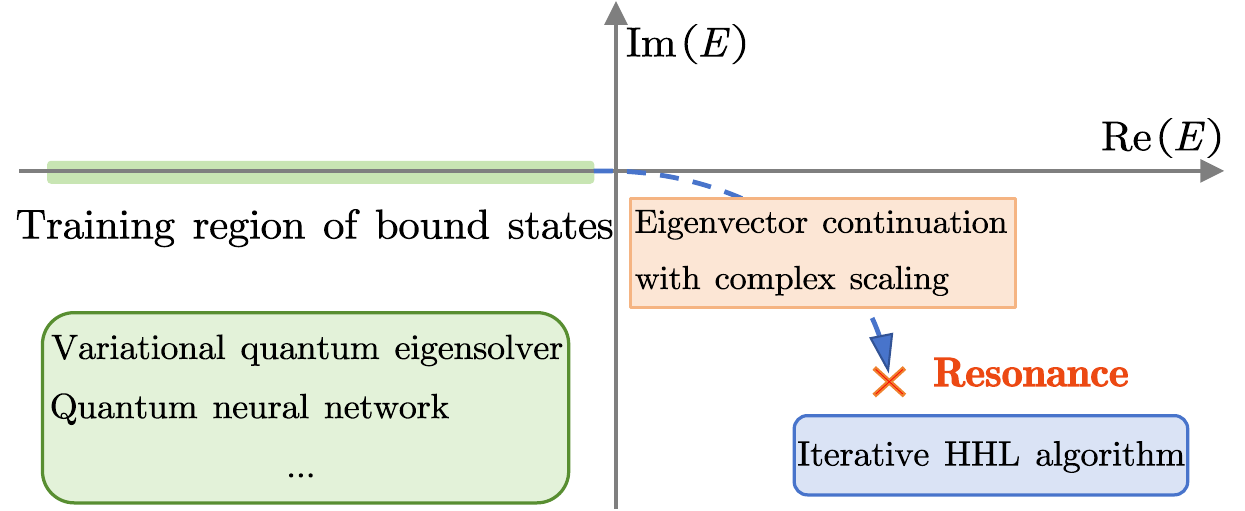} }
 	\caption{Schematic of eigenvector continuation with complex scaling. The horizontal and vertical axes represent the real and imaginary parts of energy, respectively. The bound states along the negative real axis serve as training data. After obtaining the corresponding wave functions, with eigenvector continuation complex scaled Hamiltonian (or complex scaled wave functions) will lead to the resonant energies in the fourth quadrant. The bound states can be computed using variational quantum eigensolver, quantum neural network  and so on, while the eigenvalues of the non-Hermitian operator are solved using our proposed iterative HHL algorithm.}	\label{CSEC} 
 \end{figure}

 As shown in Fig.2, the task of  bound state solutions in the training region can be handled by quantum algorithms such as the variational quantum eigensolver (VQE), quantum neural networks (QNNs) and so on. For evaluating the complex eigenvalues and eigenvectors of resonances we employ the iterative HHL algorithm illustrated in Fig.1.

\section{numerical results}
\label{Numerical Results}

As a numerical test, we consider calculating the resonant state of the $\alpha-\alpha$ system. The interaction between two $\alpha$ particles is modeled by a simple Gaussian potential,
%\begin{equation}
%	\begin{aligned}\label{}
%		&V_{\alpha-nucleon}(r)=
%		\begin{cases}
%			&(1+\beta \bm{l}\cdot \bm{\sigma})V_0, \ \ \ r<a\\&
%			0,\ \ \ r>a
%		\end{cases}
%	\end{aligned}
%\end{equation}
\begin{equation}
	\begin{aligned}\label{}
		&V_{\alpha-\alpha}(r)=V_0\exp(-\dfrac{r^2}{a^2}),
	\end{aligned}
\end{equation}
where $V_0$ is -122.6225 MeV, $a$=2.132 fm  \cite{BUCK1977246}.
 $\lambda$ parameter in total Hamiltonian $H=T+ \lambda V_{\alpha-\alpha}(r)+\frac{4e^2}{r}$ is treated as the continuation parameter and the target parameter is $\lambda_{\odot}=1$. $\lambda_{\odot}=1$ supports a $G$-wave resonant state with energy $11.8079-1.8085i$ MeV, which is computed by using $R$-matrix method. Subsequent calculations are all centered around this resonant state. 
 
First, for eigenvector continuation with complex scaling (EC with CSM), we choose a parameter range that ensures the system is bound, allowing us to obtain several ground-state wave functions 
$\phi_T=\{\phi_i, i=1,2\cdots\, i_{max}\}$. Since our focus is on solving resonant states, we simplify this process here by using traditional methods to directly obtain the ground-state wavefunctions  $\phi_T$ without quantum computing. It is important to note that the wave functions  $\phi_T$ must be orthogonalized to be  $\phi_T^O=\{\phi_i^O, i=1,2\cdots,i_{max}\}$ before constructing the complex symmetric matrix for resonant states. The complex scaling angle is set to 20 degrees and  the matrix elements of $C$ can be written as,
\begin{equation}
	\begin{aligned}\label{}
		&C_{ij}(E,1)=(\phi^O_i|{e^{-2i\theta}T+V_N(re^{i\theta})+e^{-i\theta}V_C-(E-1 )}|\phi^O_j).
	\end{aligned}
\end{equation}

\begin{figure}[htbp] 
	\centering
		{\includegraphics[width=0.4\textwidth]{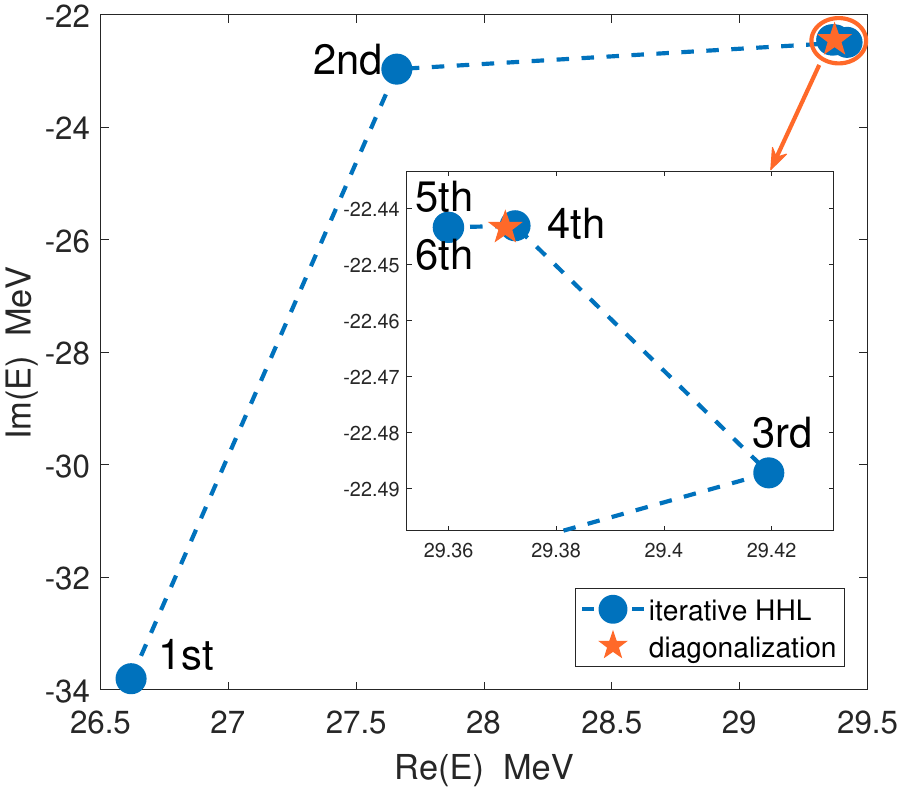} }
	\caption{The first eigenvalue obtained using the iterative HHL algorithm. After 6 iterations, the result reaches the specified precision $\varepsilon=10^{-4}$ MeV.}	\label{IHHL_first} 
\end{figure}

Here, we use total 8 training parameters uniformly distributed in the interval [1.45, 1.75] to obtain the corresponding wave functions of ground states. After orthonormalizing the basis vectors, the complex scaled Hamiltonian matrix $H^{\theta=20^{\circ}}=e^{-2i\theta}T+V_N(re^{i\theta})+e^{-i\theta}V_C$ can be obtained (listed in the appendix). In order to separate the resonant state from the continuum spectrum, the complex scaling angle generally needs to satisfy $2\theta>\arctan(\frac{Im(E_{res})}{Re(E_{res})})$.  It should be noticed that, in general when using the complex scaling method it is necessary to compute the eigenenergies corresponding to different complex scaling angles $\theta$ to obtain the final resonant energy $E_{res}$ by such stabilization condition. However, in numerical results, we only take one complex scaling angle $\theta=20^{\circ}$ as an example. In addition, considering the uncertainties in the training data, if we want to obtain more detailed results we need to analyze the error in the selection of bound states in the training region.

\begin{figure}[htbp] 
	\centering
	{\includegraphics[width=0.4\textwidth]{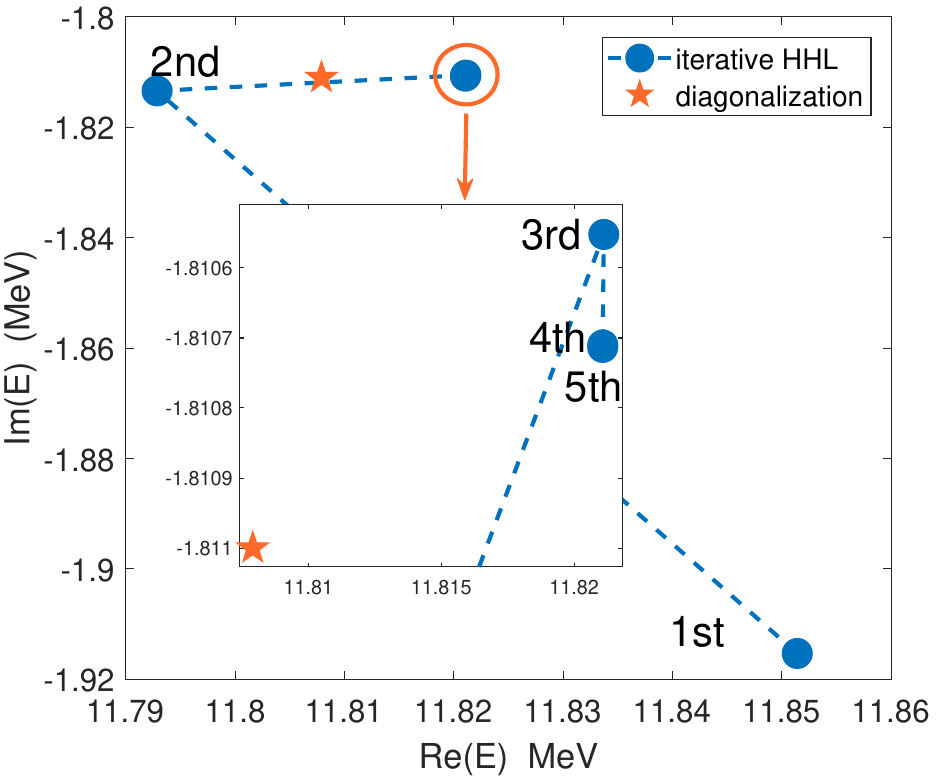} }
	\caption{The second eigenvalue obtained using the iterative HHL algorithm and projection. After 5 iterations, the result reaches the specified precision $\varepsilon=10^{-4}$ MeV. This eigenvalue represents the resonant energy.}	\label{IHHL_second} 
\end{figure}
%29.3599-22.4433i
%11.8211-1.8107i
%17.35254668134818-12.678789526633114j
%48.81633869810559-52.261985927672704j
%97.9506035631371-123.97433855270228j
%3.6297054935025876-2.7682874837854166j
%7.625438224861472-5.979003908195395j
%1.5967819041609954-1.1574359086413053j

%97.98 -              123.9785i
%48.8284 -               52.2619i
%29.3705 -               22.4434i
%17.3572 -               12.6788i
%11.8079 -                 1.811i
%7.626 -                 5.979i
%3.6284 -                2.7683i
%1.5967 -                1.1574i
\renewcommand{\arraystretch}{1.5} 
\begin{table*} \small
	\caption{The eigenenergies of $H^{\theta=20^{\circ}}(\lambda_{\odot}=1)$ obtained by iterative HHL algorithm and direct diagonalization. The energies are listed in ascending order according to the real part. The energies corresponding to $\alpha-\alpha$ $G$-wave resonant state are underlined. By $R$-matrix method the resonant energy is determined to be $11.8079-1.8085i$ MeV.}
	\begin{tabular}{ccccccccc}
		\hline
		Approach&&&&Real part of eigenenergy(MeV)&&&&\\
		\midrule
		Iterative HHL&1.5968& 3.6297&7.6254&\underline{11.8211}&17.3525&
		29.3599&48.8163&97.9506\\
	    Diagonalization&1.5967& 3.6284&7.6260&\underline{11.8079}&17.3572&
	    29.3705&48.8284&97.9800\\
		\hline
	    Approach&&&&Imaginary part of eigenenergy(MeV)&&&&\\
		\midrule
		Iterative HHL&-1.1574& -2.7683&-5.9790&\underline{-1.8107}&-12.6788&
		-22.4433&-52.2620&-123.9743\\
		Diagonalization&-1.1574& -2.7683&-5.9790&\underline{-1.8110}&-12.6788&
		-22.4434&-52.2619&-123.9785\\
		\hline
	\end{tabular}
\end{table*}
In the quantum simulation we use the HHL algorithmic interface provided by the pyqpanda of Origin's quantum \cite{OriginalQuantum}, which allows for the input of the Hermitian matrix $A$ and the real vector $b$. The interface provides different precision values representing the count of digits after the decimal point. Its default value is 0, which means that only integer solutions are available. The higher the precision, the greater the count of quantum bits and the depth of the circuit, e.g., for a precision of 1 bit, 4 additional quantum bits are needed, for a precision of 2 bits, 7 additional quantum bits are needed, and so on. In our quantum simulation we use 1-bit precision to achieve all calculations. 

The first eigenenergy and the second eigenenergy obtained using the iterative HHL algorithm are shown in Fig. 3 and 4, respectively.  
Blue circles represent the results obtained with iterative HHL and the eigenvalues obtained by direct diagonalization of the matrix $H ^{\theta=20^{\circ}}$ are marked with red asterisks. After six and five iterations respectively the first and second eigenvalues converge to our given accuracy $\varepsilon=10^{-4}$ MeV.  It can be obviously shown that using iterative HHL algorithm the converged eigenenergies will be obtained after a few iterations, which demonstrates the high computational efficiency of our algorithm. The second eigenenergy is solved using a projection method to ensure orthogonality and thus avoid convergence of the results to the first eigenstate. However, it should be noted that the order of the eigenvalue solution may change if different initial wave functions are used. The initial wave function we chose here (listed in the appendix) is such that the resonant state is already obtained in the second solution, otherwise we need to continue to use iterative HHL to complete the remaining eigenvalues until the resonant state energy is obtained. The projection strategy we utilize is: given a random initial wave function (real or complex) when solving for the $i-$th ($i > 1$) eigenvalue, and then projecting it so that it is perpendicular to the first $i-1$ eigenvectors that have already been solved, and inputting this orthogonalized initial wave function and the corresponding expectation energy into the iterative HHL algorithm. Eventually we tabulate all the eigenvalues (in ascending order of real parts) obtained from the iterative solution and give the results of the direct diagonalization as a comparison. There is a good agreement between the results obtained by the iterative HHL algorithm and those obtained by direct diagonalization as a benchmark, which validates the reliability of our algorithm for the computation of the eigenvalues of non-Hermitian matrices.

\section{conclusion}
\label{Conclusions}

In this work, we propose a novel quantum algorithm for solving resonant states based on the iterative HHL algorithm and eigenvector continuation. By combining eigenvector continuation with the complex scaling method, we can extend from bound states to resonant states. The characteristic of eigenvector continuation to reduce the solution space will help us reduce the number of qubits required and obtain more reliable results in quantum computing. To address the non-Hermiticity introduced by complex scaling, we expand the non-Hermitian matrix into a Hermitian one and propose an iterative algorithm based on HHL quantum algorithm to determine the wave function and complex energy of the resonant states. Taking the $G$-wave resonance of the $\alpha-\alpha$ system as an example, using quantum simulations we obtain the resonant energy consistent with the conventional method. We demonstrate the feasibility and reliability of our  novel non-Hermitian quantum eigensolver, which provides new algorithm and  perspective for the future study of nuclear resonance within the framework of quantum computing.

\section*{Acknowledgements}
	This work is supported by the National Natural Science Foundation of China (Grants No.\ 12035011, No.\ 11905103, No.\ 11947211, No.\ 11761161001 , No.\ 11961141003, No.\ 12022517, No.\ 12375122 and No.\ 12147101), by the National Key R\&D Program of China (Contracts No.\ 2023YFA1606503), by the Science and Technology Development Fund of Macau (Grants No.\ 0048/2020/A1 and No.\ 008/2017/AFJ), by the Fundamental Research Funds for the Central Universities (Grant No.\ 22120210138 and No.\ 22120200101).

\appendix
\onecolumn

\section{}

The appendix provides the complex-scaled Hamiltonian $H^{\theta=20^{\circ}}$ used in quantum simulations and the initial wave functions $\phi^{1st}_0$ and $\phi^{2nd}_0$ employed for calculating the first two eigenvalues,
{
\begin{equation}
	\begin{aligned}
		&Re(H^{\theta=20^{\circ}})=\\&
		\begin{bmatrix}
			5.9160 & -1.9968 & -1.4539 & 0.0808 & -0.8477 & 1.0134 & 0.6587 & 0.2180 \\ 
			-1.9968 & 20.9328 & 8.9038 & -2.0352 & 0.5117 & -0.2871 & 0.7430 & 1.2106 \\ 
			-1.4539 & 8.9038 & 20.3077 & -14.6338 & -7.1324 & 2.1843 & 0.1511 & 0.1616 \\ 
			0.0808 & -2.0352 & -14.6338 & 23.7568 & 18.6313 & -9.6136 & -3.5157 & -0.7455 \\ 
			-0.8477 & 0.5117 & -7.1324 & 18.6313 & 28.2676 & -22.9362 & -12.3736 & -5.0318 \\ 
			1.0134 & -0.2871 & 2.1843 & -9.6136 & -22.9362 & 33.7226 & 27.6980 & 15.2191 \\ 
			0.6587 & 0.7430 & 0.1511 & -3.5157 & -12.3736 & 27.6980 & 39.5660 & 32.5994 \\ 
			0.2180 & 1.2106 & 0.1616 & -0.7455 & -5.0318 & 15.2191 & 32.5994 & 45.7255 
		\end{bmatrix}, \\
		&Im(H^{\theta=20^{\circ}})=\\&
		\begin{bmatrix}
			-7.1245 & -13.5742 & -10.3647 & 6.9730 & 3.7763 & -1.3817 & -0.0922 & 0.2918 \\ 
			-13.5742 & -21.0051 & -14.8535 & 12.0314 & 9.9754 & -7.3522 & -4.2786 & -1.5986 \\ 
			-10.3647 & -14.8535 & -22.4724 & 18.8352 & 14.0059 & -10.3900 & -7.4947 & -4.8701 \\ 
			6.9730 & 12.0314 & 18.8352 & -25.0279 & -21.8145 & 15.8726 & 11.4187 & 7.9816 \\ 
			3.7763 & 9.9754 & 14.0059 & -21.8145 & -29.2031 & 25.8374 & 18.3401 & 12.6511 \\ 
			-1.3817 & -7.3522 & -10.3900 & 15.8726 & 25.8374 & -34.1859 & -30.0201 & -20.7983 \\ 
			-0.0922 & -4.2786 & -7.4947 & 11.4187 & 18.3401 & -30.0201 & -39.3154 & -34.3378 \\ 
			0.2918 & -1.5986 & -4.8701 & 7.9816 & 12.6511 & -20.7983 & -34.3378 & -44.7440 
		\end{bmatrix}
	\end{aligned}
\end{equation}
\begin{equation}
	\begin{aligned}
\phi^{1st}_0 =[
	-0.50298,\ -0.10005,\ -0.17812,\ -0.4794,\ 0.74079,\ -0.62992,\ -0.96068,\ 0.9065
]^T,
	\end{aligned}
\end{equation}
\begin{equation}
	\begin{aligned}
\phi^{2nd}_0 =[ &0.0799 - 0.4118i,\ 0.3987 + 0.0799i,\ -0.6690 - 0.0337i,\ -0.0204 - 0.0322i,\\ &\ -0.1536 + 0.0933i,\ -0.0851 - 0.1819i,\ -0.1936 + 0.1947i,\ 0.2308 + 0.0854i ]^T.
	\end{aligned}
\end{equation}
}

The eigenenergies that converge after iteration are $29.3599-22.4433i$ and $11.8211-1.8107i$ respectively, and the corresponding normalized eigenvectors are listed int he following,
\begin{equation}
	\begin{aligned}
\phi^{1st}=[  &-0.2610 - 0.1947i,\   0.0084 - 0.5524i,\ 
0.0919 - 0.3528i,\   -0.1403 - 0.0176i,\ \\
&-0.1529 - 0.3527i,\    0.0848 + 0.3638i,\ 
-0.0487 + 0.0000i,\   -0.1337 - 0.3581i]^T\\
\phi^{2nd,res}=[   &0.1228 - 0.8607i,\   0.1846 + 0.3683i,\
0.1756 + 0.1208i,\   0.0321 + 0.0417i,\ \\
&0.1337 + 0.0011i,\  -0.0351 + 0.0521i,\
0.0420 - 0.0007i,\  -0.0122 - 0.0499i]^T.
	\end{aligned}
\end{equation}

Directly diagonalizing the Hamiltonian $H^{\theta=20^{\circ}}$  yields the eigenvalues as follows,
\begin{equation}
	\begin{aligned}
E_{diag}=[&97.9800 -123.9785i,\ 
48.8284 -52.2619i,\ 
29.3705 -22.4434i,\ 
17.3572 -12.6788i,\\
&11.8079 -1.8110i,\ 
7.6260 -5.9790i,\ 
3.6284 -2.7683i,\ 
1.5967 -1.1574i].
\end{aligned}
\end{equation}

\twocolumn

\bibliographystyle{elsarticle-num-names} 
%\nocite{*}
\bibliography{example}

\end{document}